\documentstyle[aps,epsfig]{revtex}
\begin{document}
\title{\Large\bf {A parameter-free optical potential for the heavy-ion 
elastic scattering process}}
\author{
M. A. G. Alvarez, L. C. Chamon, M. S. Hussein, D. Pereira, L. R. Gasques, 
E. S. Rossi Jr., C. P. Silva.}
\address{ 
Departamento de F\'{\i}sica Nuclear, Instituto de F\'{\i}sica
da Universidade de S\~{a}o Paulo,\\
Caixa Postal 66318, 05315-970, S\~{a}o Paulo, SP, Brazil.}
\maketitle
\begin{abstract}
Thirty elastic scattering angular distributions for seven heavy-ion systems, 
in wide energy ranges, have been studied with the aim of systematizing 
the optical potential, real and imaginary parts, in a global way. The 
framework is: i) an extensive systematization of nuclear densities, ii) the 
energy dependence of the bare potential accounted by a model based on the 
nonlocal nature of the interaction, and iii) the real and imaginary parts of 
the optical potential assumed to have the same radial shape.
\end{abstract}

\vspace{5mm}


\vspace{5mm}

Elastic scattering is the simplest and most direct process involved in a 
nuclear reaction, and it can be used as the starting point to understand more 
complicated reaction channels. Over the last decades, different models were 
used for the real and imaginary parts of the optical potential to reproduce a 
large number of elastic scattering data involving heavy-ion systems. The 
optical model analysis using the conventional Woods-Saxon (WS) shape for the 
real and imaginary parts of the potential, although far from being fundamental, 
has presented the best means for the reproduction of the elastic scattering 
angular distributions, with six free parameters used in the data fits. However, 
there are problems in terms of obtaining a simple model for systematizing the 
WS optical potential parameters, in order to take into account the energy 
dependence, refractive effects in light heavy-ion systems, exotic nuclei 
systems etc. Particularly, for some systems the variation with the bombarding 
energy requires arbitrarily different sets of parameters, with different 
shapes and strengths for the potential. This arbitrariness calls for a more 
realistic model for the optical potential, which has been accomplished to a 
large extent by our work \cite{1,2,3,4,14} on the nonlocal model for the real 
part of 
the nucleus-nucleus interaction. The central idea of the present work is to 
perform a further test of consistency of this model for the real part of the 
interaction, by using a very simple form to describe the imaginary part of the 
optical potential, and avoiding as much as possible the use of free 
parameters in accounting to the data.

A previous investigation \cite{7,8,9} to identify similarities between the real 
and imaginary parts of the potential has been performed by considering the 
realistic Lax-type interaction, which has provided satisfactory fits of elastic 
scattering data at intermediate energies \cite{7,8}. The Lax interaction 
\cite{Hu} (Eq. 1) is the optical limit of the Glauber high-energy-approximation 
\cite{5,6}, and it is essentially the zero-range double-folding potential used 
for both the real and imaginary parts:
\begin{equation}
U(R)=-\frac{1}{2}\hbar v \int (\alpha +i) \; \sigma_T^{NN} \; \rho_T 
(\vec{r'}) \; \rho _P (\vec{R}-\vec{r'}) \; d\vec{r'}
\end{equation}
where $v$ is the relative velocity between the nuclei, $\sigma^{NN}_T$ is a 
spin-isospin-averaged total nucleon-nucleon cross section, $\rho_P$ and 
$\rho_T$ are the projectile and target nuclear densities, and $\alpha$ is a 
known energy-dependent quantity that determines the real part of the 
nucleon-nucleon forward elastic amplitude $f_{NN}(E,0)$. Eq. (1) is obtained 
from the optical theorem applied to $f_{NN}(E,0)$. The Lax-type interaction 
is not valid for low energies where collective reaction processes are 
important. Thus, the parameter-free description of low energy data is an open 
question in the determination of a fundamental potential. However, the 
procedure of using the same radial shape for the real and the imaginary parts 
of the potential has successfully been used in the present work.

In \cite{1,2,3,4,14}, we have developed another realistic model for the 
heavy-ion bare interaction, which takes into account the Pauli nonlocality 
involving the exchange of nucleons between the target and projectile. This 
model has presented the same validity at low and high energies, and has 
already been tested for a large number of systems \cite{1,2,3,14,12,15,16,17}. 
Within the nonlocal model, the bare interaction $V_N$ is connected with the 
folding potential $V_F$ through \cite{14}
\begin{equation}
V_N(R,E)\approx V_F(R)\,e^{-4v^2/c^2}
\end{equation}
where c is the speed of light and $v$ is the local relative velocity between 
the two nuclei,
\begin{equation}
v^2(R,E)=\frac{2}{\mu}[E-V_C(R)-V_N(R;E)]
\end{equation}
The folding potential (Eq. 4) can be obtained in two different ways 
\cite{14}: i) using the nucleon distributions of the nuclei and an
appropriate form for the nucleon-nucleon interaction, and ii) using the
matter distributions of the nuclei with a zero-range approach for
$v(\vec{r})$. By matter distribution we mean taking into account the finite 
size of the nucleon. Both alternatives are equivalent in describing the 
heavy-ion nuclear potential \cite{14}, and in the present work we have 
adopted the zero-range approach.
\begin{equation}
V_F(R) = \int \rho_1(r_1) \; \rho_2(r_2) \; v(\vec{R}-\vec{r_{1}}+\vec{r_{2}})
\; d\vec{r_1} \; d\vec{r_2} \;
\end{equation}
For the Coulomb interaction, $V_C$, we have used the expression for the 
double-sharp cutoff Coulomb potential \cite{18}. This procedure is important 
in calculating cross sections at intermediate energies, where the internal 
region of the interaction is probed.

With the aim of providing a global description of the nuclear interaction, a 
systematization of nuclear densities has been proposed in Ref. \cite{14}, 
based on an extensive study involving charge distributions extracted from 
electron scattering experiments and theoretical densities calculated through 
the Dirac-Hartree-Bogoliubov model. In that study, we have adopted the 
two-parameter Fermi (2pF) distribution to describe the nuclear densities. The 
radii of the 2pF distributions are well described by
\begin{equation}
R_{0} = 1.31A^{1/3} - 0.84\; fm ,
\end{equation}
where $A$ is the number of nucleons of the nucleus. The matter densities 
present an average diffuseness value $a =0.56 \; fm$. Owing to specific nuclear 
structure effects (single particle and/or collective), the parameters $R_0$ 
and $a$ show small variations around the corresponding average values 
throughout the periodic table. However, as far as 
the nuclear potential is concerned, the effects of the structure of the nuclei 
are mostly present at the surface and mainly related only to the diffuseness 
parameter \cite{14}. This systematization of the nuclear distributions is 
essential to obtain a parameter-free interaction, since the folding potential 
depends on the densities of the partners in the collision. Within this context, 
an extensive systematization of optical potential strengths extracted from 
heavy-ion elastic scattering data analyses at low and intermediate energies 
was performed \cite{14}, and the experimental potential strengths have been 
described within $25\%$ precision.

As mentioned above, an important point that stands out in obtaining a 
description of the optical potential in a global way is the difficulty 
encountered in describing the imaginary part of the interaction within a simple 
model. A fully microscopic description based on the Feshbach theory is very 
difficult, and basically out of reach at low energies where collective 
as well as single particle excitations are involved in the scattering process. 
In previous works involving elastic scattering data fits 
\cite{2,3,12,15,16,17}, we have 
already used the nonlocal model for the real part of the interaction, and 
adopted a more modest procedure for the imaginary part by assuming two 
different models: WS with three free parameters, which has presented an 
excellent description of the data; and the parameter-free Lax-type 
approximation, which is based on a more fundamental theory but has not been 
used to describe low energy data. Motivated by the concept from the Lax 
approximation of using similar shapes for the real and imaginary parts of the 
potential, in this work we have extended Eq. (2), developed for the real part 
of the interaction, to the imaginary part of the potential, by simply 
multiplying it by $N_i$, where $N_i$ is a number to be fixed by adjusting the 
data.

\begin{equation}
W(R,E)= N_i \; V_N(R,E)
\end{equation}.

We have chosen the $^{12}$C + $^{12}$C, $^{16}$O, $^{40}$Ca, $^{90}$Zr, 
$^{208}$Pb; $^{16}$O + $^{208}$Pb and $^{40}$Ar + $^{208}$Pb 
systems as test cases due to the extensive experimental data available 
\cite{19,20,21,22,23,24,25,26,27,28} over wide energy ranges, and 
principally because the special refractive characteristics involving some of 
these systems which makes them more sensitive to the real part of the 
interaction. In Figures (1) to (7) the solid lines correspond to the best data 
fits obtained by searching the $N_i$ parameter. We have opted for keeping the 
average density diffuseness value $a=0.56\,fm$ in the calculations, even 
though we could improve the quality of the data fits by allowing the 
diffuseness to be a free parameter. 
As one can observe (see Fig. 8), the $N_i$ parameter is approximately 
system-independent, with an average value $N_i=0.78$. Good elastic scattering 
cross section predictions are obtained using this average value for the whole 
data set (see the dotted lines in Figs. 1 to 7).

In summary, using the procedure described above we have obtained a good 
description of the whole data set, which has further validated our assumption 
for the real part of the interaction: the nonlocal model. We have also assumed 
a very simple model for the imaginary part of the potential, with only one, 
system- and energy-independent, free parameter: the average value for 
$N_i \approx 0.8$. In fact, the details of the imaginary part of the 
interaction seems not to be of much importance to the data fit. For example, 
in Fig. 7 (bottom) quite different values for $N_i$ $(N_i=0.44$ and 
$N_i=0.78)$ provide very similar predictions for the elastic scattering cross 
sections. The same behavior can be observed for the other systems (see Figs. 1 
to 7). At a first glance this result seems to be surprising, but upon a 
second thought one does expect that the data are more sensitive to the real 
part of the potential, which determines the quantal transmission through the 
$\ell$-dependent barriers. In order to confirm this point, we have compared 
(see table 1) the reaction cross sections resulting from our optical model 
(OM) calculations (using the average value for $N_i$) with those from the 
geometrical limit of the barrier penetration model (Eq. 7) 
\begin{equation}
\sigma_{BP}= \pi R_B^2 \; (1 - V_B/E)
\end{equation}
In most cases both forms of calculating the reaction cross section agree within 
about 20\% precision. However, the values obtained from the OM calculations 
are more realistic. Indeed the reaction cross section values obtained with 
our OM calculations are very similar (see table 1) to those obtained through 
different methods in earlier works \cite{7,19,20,24,25,26,29,30,31,32,33,34}. 
Our results suggest using the present parameter-free model to get reliable 
estimates for heavy-ion elastic scattering and reaction cross sections. 
Extension of our findings to halo-nuclei is being pursued and will be 
presented elsewhere.

This work was partially supported by Financiadora de Estudos e Projetos 
(FINEP), Conselho Nacional de Desenvolvimento Cient\'{\i}fico e 
Tecnol\'{o}gico (CNPq) and Funda\c{c}\~{a}o de Amparo \`{a} Pesquisa do 
Estado de S\~{a}o Paulo (FAPESP), under contract number 1998/11401-4.

\newpage

\indent {\bf Table 1:} The optical model ($\sigma_{OM}$) and geometrical 
($\sigma_{BP}$) reaction cross sections obtained in this work for 
several systems and bombarding energies. The table also presents the values 
(and corresponding references) for the reaction cross sections ($\sigma_R$) 
obtained through different methods in earlier works.
\begin{center}
\begin{tabular}{c|c|c|c|c|c} 
System&$E_{Lab}$ (MeV)&$\sigma_{OM}$ (mb)&$\sigma_{BP}$ (mb)&$\sigma_{R}$ (mb) 
& Ref. \\ \hline
&16& 542&508&600&30 \\
&112& 1477&1759&1444&32 \\
$^{12}$C + $^{12}$C&300& 1374&1834&1296&31 \\
&1016& 1055&1680&1000&24 \\
&1449& 910&1560&907&24 \\ \hline
&21& 209&167&-&- \\
$^{16}$O + $^{12}$C&260& 1560&1877&1481&18 \\
&608& 1420&1881&1374&18 \\
&1503& 1143&1716&1136&23 \\ \hline
&180& 2022&2083&2165&22 \\
$^{12}$C + $^{40}$Ca&300& 2015&2168&2030&22 \\
&420& 1969&2180&2000&22 \\ \hline
&120& 2310&2113&2219&22 \\
$^{12}$C + $^{90}$Zr&180& 2540&2406&2297&22 \\
&300& 2652&2610&2415&22 \\
&420& 2650&2670&2840&22 \\ \hline
&96& 1854&1498&1791&28 \\
&116& 2323&1944&2235&28 \\
&300& 3205&3013&3300&22 \\
$^{12}$C + $^{208}$Pb&420& 3611&3393&3561&22 \\
&480& 3634&3441&-&- \\
&1030& 3560&3501&-&- \\
&1449& 3344&3331&3136& 6 \\
&2400& 2905&2882&2960&27 \\ \hline
&129.5& 1993&1553&2023&28 \\
$^{16}$O + $^{208}$Pb&192& 2878&2408&2847&28 \\
&312.6& 3504&3067&3432&29 \\
&1500& 3708&3530&3485&23 \\ \hline
$^{40}$Ar + $^{208}$Pb&302& 2476&1740&-&- \\
&1760& 4783&4059&-&- \\ \hline
\end{tabular}
                                
\end{center}

\newpage

\begin{figure}
\vspace*{14.0cm}
\hspace{2.0cm}
\includegraphics{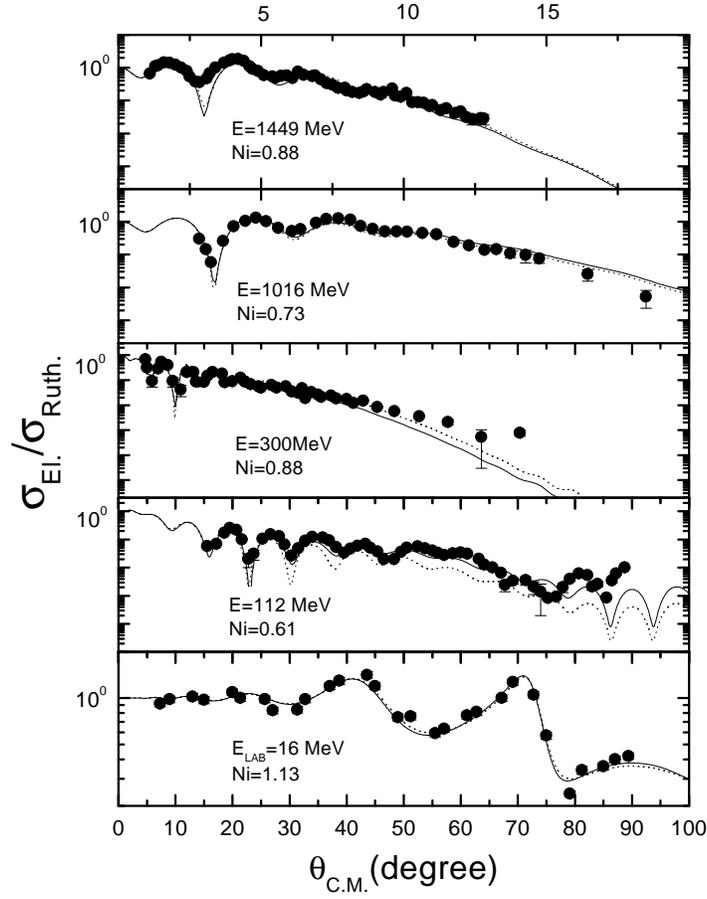}
\vspace{1.0cm}
\noindent
\caption{Elastic scattering angular distributions for the $^{12}$C + 
$^{12}$C system in several bombarding energies. The solid lines correspond to 
the best fit using the same radial shape for both the real and imaginary parts 
of the optical potential, with the $N_i$ parameter searched for the best data
fits. The dotted lines correspond to the predictions obtained with the average 
value $N_i=0.78$ (see text for details).} 
\end{figure}

\newpage

\begin{figure}
\vspace*{14.0cm}
\hspace{2.0cm}
\includegraphics{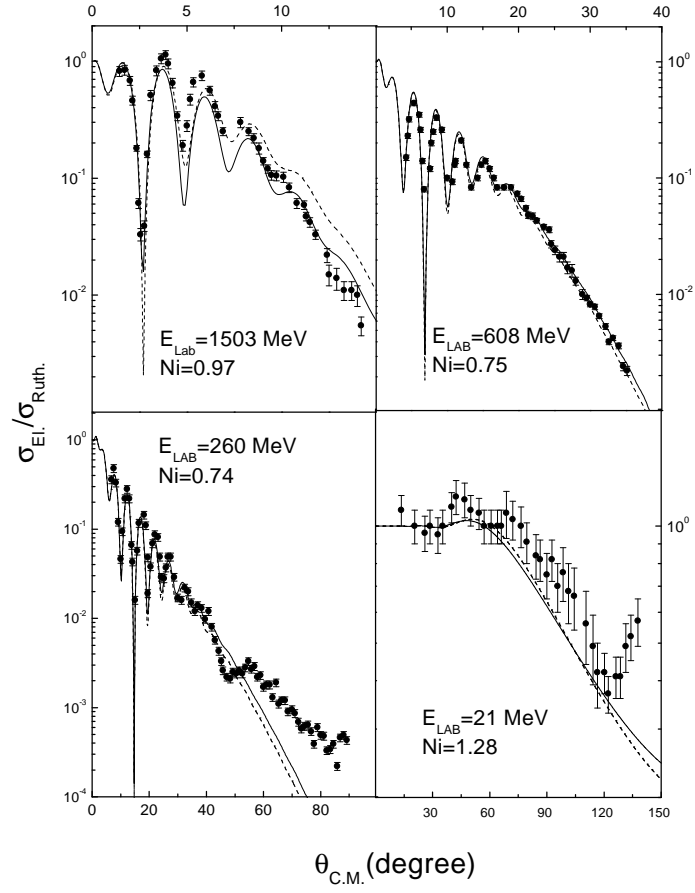}
\vspace{1.0cm}
\noindent
\caption{The same of Fig. 1 for the $^{16}$O + $^{12}$C system.}
\end{figure}
 
\newpage

\begin{figure}
\vspace*{14.0cm}
\hspace{2.0cm}
\includegraphics{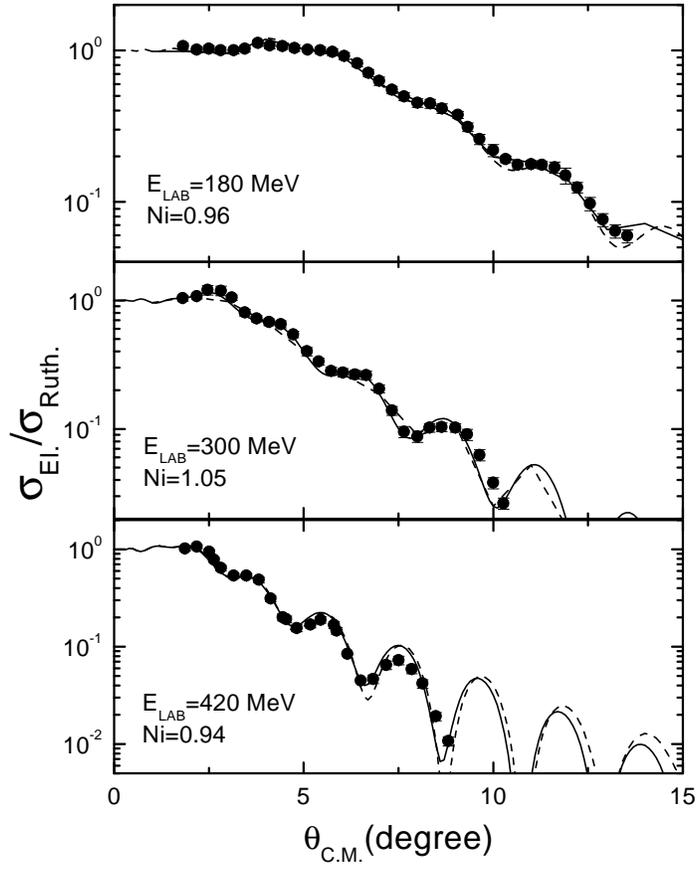}
\vspace{1.0cm}
\noindent
\caption{The same of Fig. 1 for the $^{12}$C+$^{40}$Ca system.}
\end{figure}

\newpage

\begin{figure}
\vspace*{14.0cm}
\hspace{2.0cm}
\includegraphics{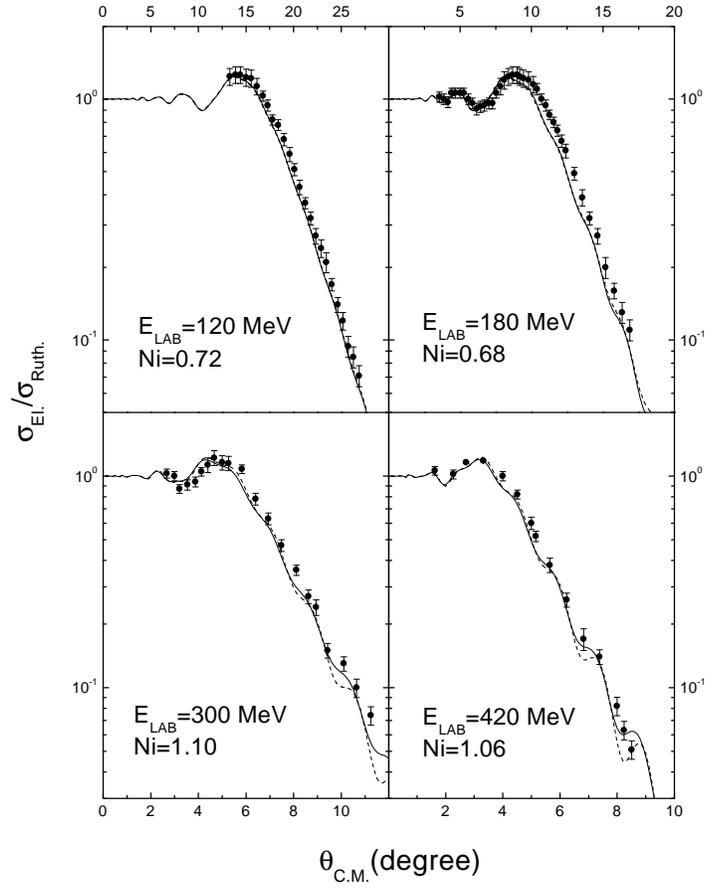}
\vspace{1.0cm}
\noindent
\caption{The same of Fig. 1 for the $^{12}$C+$^{90}$Zr system.}
\end{figure}

\newpage

\begin{figure}
\vspace*{14.0cm}
\hspace{2.0cm}
\includegraphics{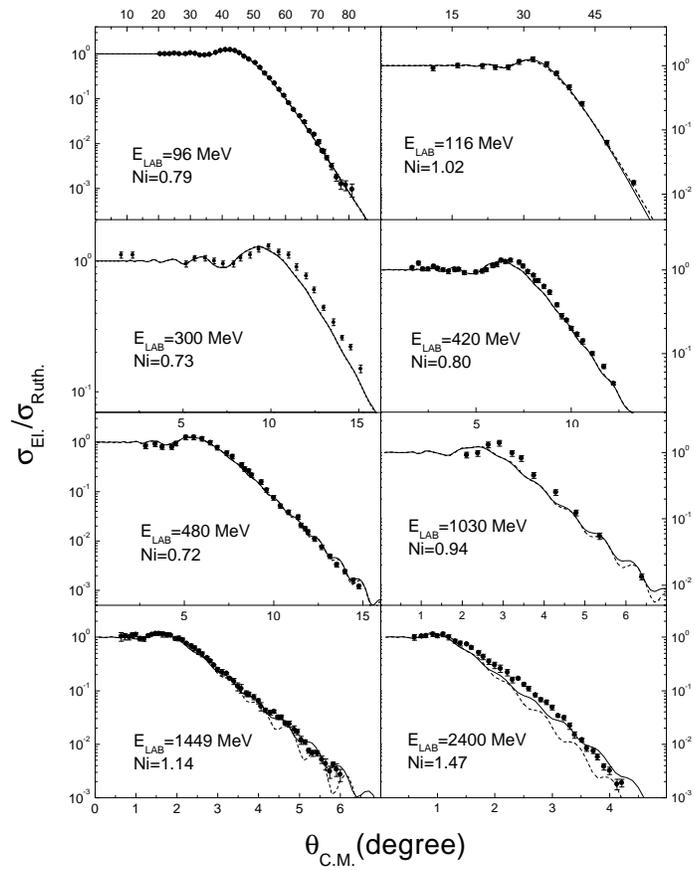}
\vspace{1.0cm}
\noindent
\caption{The same of Fig. 1 for the $^{12}$C+$^{208}$Pb system.}
\end{figure}

\newpage

\begin{figure}
\vspace*{1.0cm}
\hspace{2.0cm}
\includegraphics{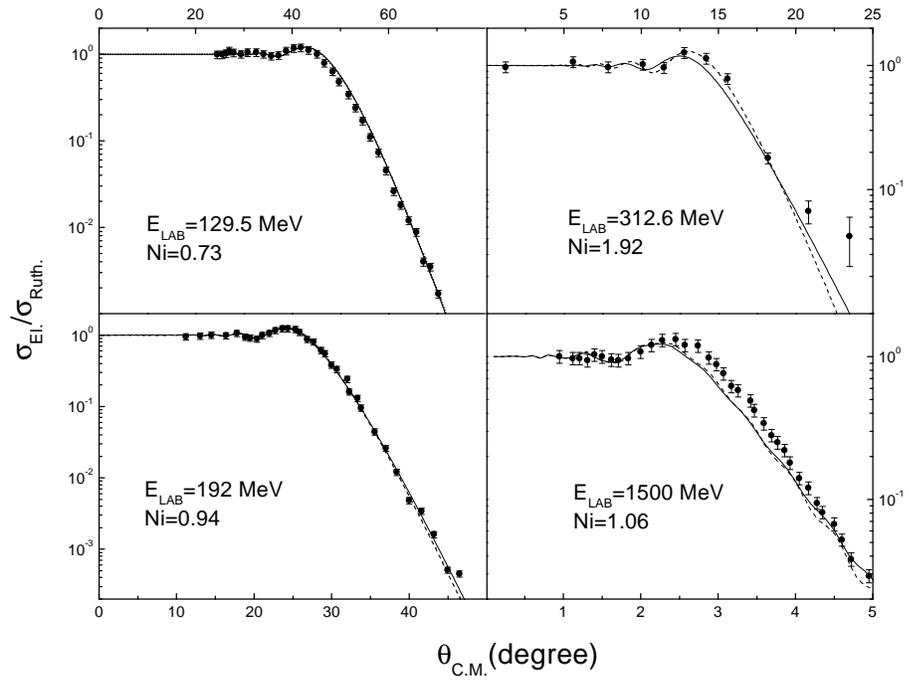}
\vspace{12.0cm}
\noindent
\caption{The same of Fig. 1 for the $^{16}$O+$^{208}$Pb system.}
\end{figure}

\newpage

\begin{figure}
\vspace*{1.0cm}
\hspace{2.0cm}
\includegraphics{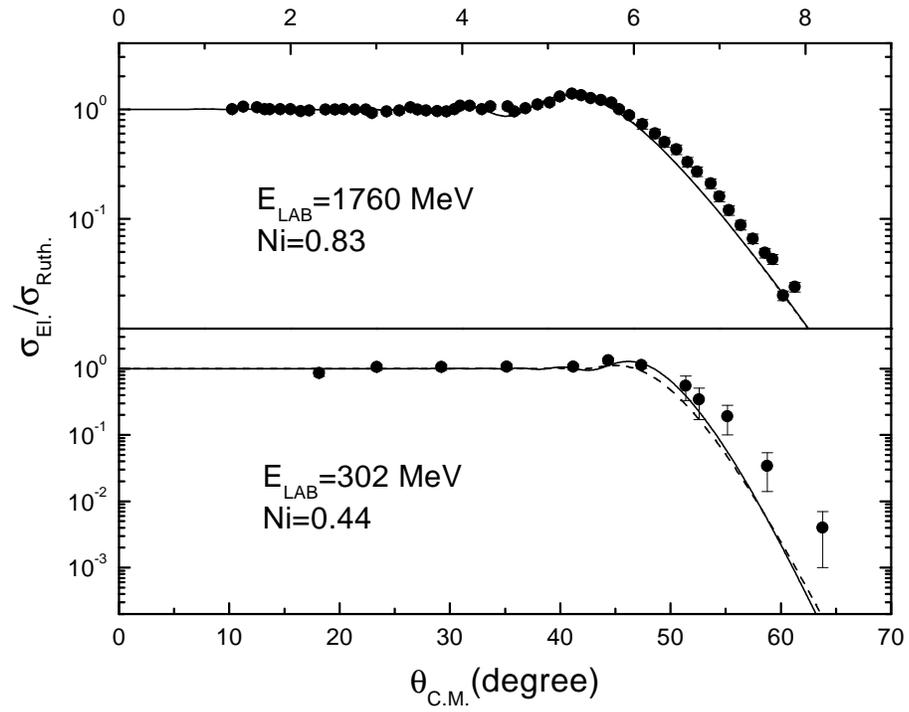}
\vspace{12.0cm}
\noindent
\caption{The same of Fig. 1 for the $^{40}$Ar+$^{208}$Pb system.}
\end{figure}

\newpage

\begin{figure}
\vspace*{1.0cm}
\hspace{2.0cm}
\includegraphics{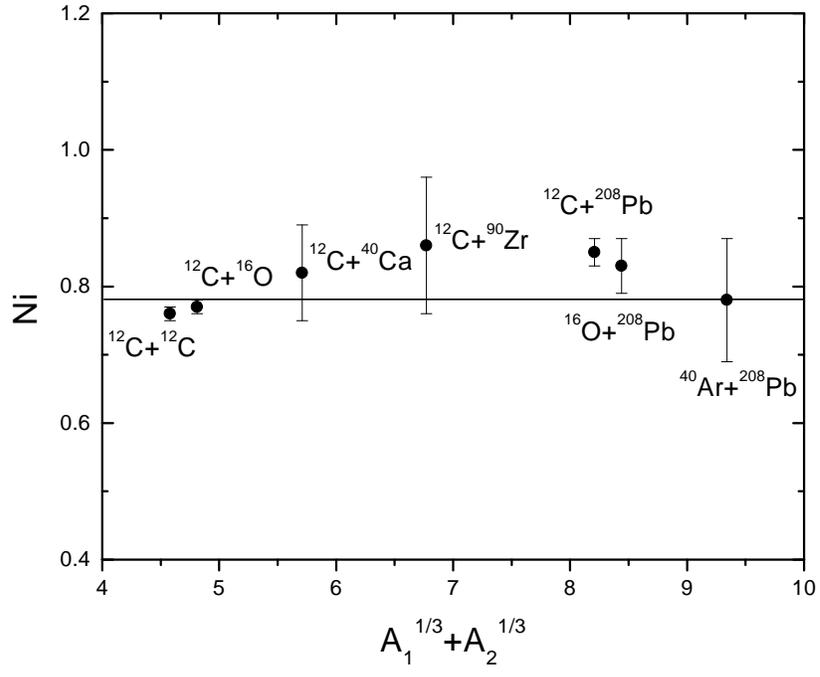}
\vspace{12.0cm}
\noindent
\caption{The $N_i$ values for different systems obtained by adjusting the 
corresponding elastic scattering angular distributions. The solid line 
represents the average value $N_i=0.78$.}
\end{figure}

\end{document}